\documentstyle[12pt,aasms4]{article}
 
\begin{document}
 
\def\etal{{\it et al.\/\ }}
\def\today{\ifcase\month
           \or January   \or February \or March    \or April
           \or May       \or June     \or July     \or August
           \or September \or October  \or November \or December\fi
           \space\number\day, \number\year}
\def\ga{\lower 2pt \hbox{$\, \buildrel {\scriptstyle >}\over{\scriptstyle \sim}\
,$}}
\def\la{\lower 2pt \hbox{$\, \buildrel {\scriptstyle <}\over{\scriptstyle \sim}\
,$}}

\title{\bf PRINCIPLES AND APPLICATIONS OF DIFFERENCE IMAGING}

\author{Austin B. Tomaney}

\affil{Department of Astronomy, University of Washington}

\centerline{\it E-mail: austin@astro.washington.edu}

\vskip 2in

\centerline{\it submitted to the Astronomical Journal, 23 January 1998}

\vskip 2in

%\centerline{\bf \today}

\begin{abstract}

The principles of difference imaging outlined and 
the technique of Alard and Lupton (1997) is generalised 
to generate the
best possible difference images to within the limits of measurement
error. It is shown how for a large database of images we can approach
diffraction-limited spatial resolution to within measurement error 
of a combined image. This is achieved through an iterative procedure 
of difference imaging
and deconvolution. The design of the ideal imager 
to approach this limit is discussed. 

\end{abstract}
 
\clearpage
 
\section{INTRODUCTION}

This work reviews a number of techniques for image processing
and suggests how they can be put together to maximise spatial resolution
in generalised imaging systems.

The field of microlensing has made tremendous progress since Paczynski's 1986
paper on the subject. This has largely driven new techniques to improve the 
detection efficiency of microlensing events ({\it e.g.,} Tomaney \& 
Crotts 1996, based on the earlier work of Ciardullo, Tamblyn \& Phillips 1990
and Phillips \& Davis 1995) which have general application to detecting 
variability of any kind in an image. Recently Alard \& Lupton (1997) 
outlined an elegant algorithm for difference imaging. Here we develop 
these ideas further.

\section{PRINCIPLES OF DIFFERENCE IMAGING}

We review the general principles of difference imaging and mention some
specifics in the context of microlensing.

1. There is no object in the sky that is not blended to
some finite extent. 

2. However, if this object varies and we perform a difference 
image analysis it can be isolated as being one or
some combination of the following things:

{\it (a)} a single star that has intrinsically varied or has 
been microlensed,

{\it (b)} a proper motion of a star on the sky. This might
be seen as a bipolar residual Point Spread Function (PSF) that has
spatially separate negative and positive components in the Difference 
Image (DI) (this effect depends on the flux ratio of the star in the 
pre-subtracted image and reference image and is largest when the
ratio is unity),

{\it (c)} an apparent astrometric shift due to microlensing 
redistributing the star's light on the sky as the lens
transverses the line of sight,

{\it (d)} a superposition of more than one star 
that is photometrically and/or astrometrically varying (because 
of any of the above or simply
the intrinsic variability of the star(s) within the 
detected PSF on the DI),

{\it (e)} a purely atmospheric effect which shifts the star's light 
in ways that depend on its spectrum (i.e. ``colour'') and
the observation taken at a given airmass and parallactic
angle (see Filippenko 1982). (The atmospheric effect, however, can be 
addressed to some extent for every pixel in the DI analysis since
we can know each pixel's colour, airmass, and parallactic angle
and can correct for the first order effect of the location of 
the object centroid in the bandpass observed in which depends 
on the spectrum as well as the second order effect associated with
the PSF shape dependence on the object's colour, airmass
and parallactic angle. However, the accuracy is limited by how much
of the pixel's light actually comes from one star and the problems
of finite bandpasses where multi-band photometry indicates
two stars which actually have different spectra are measured to have
the same apparent colour.)

{\it (f)} a purely local phenomenon: examples would be detector defects
(especially including non-linearity in detector response), 
satellite tracks and many other artificial phenomena.

Note that a residual that is significant on a DI
can be distinguished from a star by its 
consistency or otherwise by the PSF appropriate for the 
DI. This can be used as a discriminator between cosmic rays
and CCD defects in a formal manner, except a star that has 
varied during the exposure will
have a PSF that is inconsistent with the mean PSF for 
all non-varying stars in the frame (an extreme example is a cosmic
ray). However, this discrimination is only true if the star
varies very slowly during the duration of the exposure, since the
atmospheric conditions and instrument focus are constantly changing 
during the exposure. 

To the extent that many of the effects in (a) to (e) are small, a DI 
largely removes blending with other stars so the photometric effect of
apparent colour shifting in the microlensing case ({\it e.g.,} Kamionkowski 
1995) and the associated apparent astrometric centroid shift is removed.

2. Note that for the two extremes of variability: periodic
and single exclusions (such as microlensing) we can get centroids that 
are good to the accuracy of centroid measurement on a single frame, ${dx,dy}$, 
divided by $\sqrt {N_{frames}}$ in the first case and 
$\sqrt {N_{peak-frames}}$ in the second (assuming ${dx,dy}$ are the
same for every measurement) allowing for 
accurate centroiding depending on how large $N$ is in each case.

3. If we compare two bandpasses at wavelengths $\lambda_1$ and
$\lambda_2$ then,
barring finite source effects and limb darkening (Alcock {\it et al.}
1997), 
the following must be true for microlensing:

$$ {{F (\lambda_1, t) - F_o (\lambda_1, t_{baseline})} \over
{F (\lambda_2, t) - F_o (\lambda_2, t_{baseline})}} = constant, \eqno (1.1) $$

to within the limits of measurement, in
contrast to the blended case which has an extra term (the zero-point)
to be solve for when plotting $F (\lambda_1, t)$ against $F (\lambda_2, t)$.

4. We can speak only in terms of the one parameter we can 
{\it measure} in the DI Light Curve (DILC), the FWHM of the curve, $t_{fwhm}$, 
not the Einstein crossing time, $t_E$.

5. We can use 1-3 to {\it attempt} to solve for the amplitude, $A$, of the 
event with the best available high-resolution image.

6. We can use the DILC together with the DI Astrometric 
Curve (DIAC) to attempt to break the degeneracies involved and solve
for the projected Einstein radius, $\theta_E$.

7. We still have blending when we go to the ``source'' in the
baseline image to solve for $A$. It helps if 
we can get an high-resolution image as the star is being lensed to 
solve for $A$. We need to {\it measure} the Luminosity Function 
(LF) to determine an amplitude detection efficiency for a given surface 
brightness, S, unless the LF is a monotonic power law (Gould 1996). The
LF can be measured from fits to the skewness of the intensity value 
histogram of the surface brightness fluctuations (Tomaney and Crotts
1996). However, for those light curves in which we have large S/N ($>$ 100)
the timescale/amplitude degeneracy can be broken (Gould 1996).
 
\section{DIFFERENCE IMAGING}

Consider a Gedanken experiment involving an image, $r$, taken at a location 
above the atmosphere 
with a detector $D$ which is to be compared to a similar image, $i$, taken with
the same detector at the same time where the centroids of the sources imaged
in both cases are spatially coincident on the two
detectors. There must be a convolution kernel,
$k$, describing the mapping of photons on one image 
to the other (largely describing the smearing effects of the atmosphere and 
instrumental focus on the photon paths). This convolution kernel is spatially
dependent in the image plane due to atmospheric effects and abberations of 
the imager. Thus,

$$ i = r \otimes k, \eqno (3.1) $$

If the Fourier Transform (FT) of $r$, $i$ and $k$ are $R$, $I$ and $K$ 
respectively, the Convolution Theorem states,

$$ I = K R, \eqno (3.2) $$

Thus, if $K^{-1}$ is the inverse of the matrix, $K$,

$$ K^{-1} I = K^{-1} K R = R, \eqno (3.3) $$

and,

$$r = {\rm FT} (K^{-1} I), \eqno (3.4) $$

Generally,

$$ i ({\bf x'}) = [k ({\bf x}) \otimes r ({\bf x})] s ({\bf x}) + b ({\bf x}), 
\eqno (3.5) $$

where ${\bf x}$ are the reference image coordinates and ${\bf x'}$ are
the image coordinates. The transformation,

$$ {\bf dx} = {\bf x} - {\bf x'};
 {\bf dy} = {\bf y} - {\bf y'}, \eqno (3.6) $$

represents the geometric registration of the image $r$ to $i$. This
transform generally is of low order for most imaging telescopes.
Note that for a detector, $D$, that images simultaneously in two bandpasses
every pixel can be defined with a colour (bandpass flux ratio after subtraction
of the sky), then to first order (blending issues aside) this transformation
may also contain for every pixel the correction for airmass (and parallactic
angle, since airmass has a dual degeneracy in this term) to align pixels taken
at different airmasses to that above the atmosphere.

The remaining terms of equation (3.5) above are $s ({\bf x})$, the photometric
(scaling) normalisation and $b ({\bf x})$ the sky difference (zero-point
correction) and the convolution kernel, $k ({\bf x})$ to bring the reference
image to the same Point Spread Function (PSF) as the ground-based image,
$i$. All these terms are functions of ${\bf x}$, but note that like the PSF for
a given location in ${\bf x}$ the convolution kernel is infinite in spatial 
extent. 

Let us assume that our detector is a photon counting device (such as a CCD
detector) which detects photons in
spatially discrete bins (henceforth referred to as pixels). For 
simplicity we choose an array of such pixels of $N\times N$ in
size. We can now follow the technique of Alard and Lupton (1997) and 
use equation (3.5) to express the intensity of every pixel in image 
$i$ as a function of the intensity of every pixel another ground-based
image, $i'$. (Note that this approach is working directly on the raw
data of both images respectively.) If we use 
$N_{dx}, N_{dy}, N_{s}, N_{b}, {\rm and} N_{k}$ coefficients to solve
equation (3.5), then the number of elements which can be solved for in 
the $k$ matrix is given by,

$$ {N^2 \over {2 N_{dx} N_{dy} N_{s} N_{b} {N_{k}}^2}}, \eqno (3.7) $$

In the experience of this author for a 2K by 2K CCD image in a number
of wide-field telescopes all of the terms in the denominator are
quite small compared with the extent of the kernel which can be solved
for up to a maximum of 1K by 1K if all of these terms are unity. Given
the complexity of a typical PSF in an imaging system (due to diffraction
annuli from the optical components as well as diffraction spikes from 
``spiders'' holding the secondary mirror in place in a typical Cassegrain
design telescope) it would appear sensible to make few assumptions about
the actual shape of the PSF (and consequently the matching kernel, $k$). 
However, we can probably converge on a solution for $k$ more quickly by
making reasonable assumptions about the actual shape of the PSF and 
the kernel by expanding its polynomial expression with a basis function 
that best describes the gross properties of the PSF; in an astronomical
image this is typically a Gaussian (Alard and Lupton 1997).

\subsection{Algorithm for the Practical Estimation of $k$}

We can estimate $k$ to the extent that we are limited by the following
four critical aspects of measurement,

{\it 1)} The truncation effect of $N^2$ pixels on a detector which 
effectively constrains our ability to measure the infinite extent of
$k$ to $(N / 2)^2$ pixels in size.

{\it 2)} The photon and detector noise (typically the read-out noise in
a CCD, but often will include the systematic errors in the calibration
of the data).

{\it 3)} The discrete sampling of the PSF by the detector elements.

{\it 4)} A source which is spatially or photometrically variable during
the exposure.

Let us consider differencing our best spatial resolution image, $i_r$, against
all other ($n-1$) images that we have in a time-series. The steps for 
doing this are,

{\it (i)} Solve for $k_n$ for image $n$ (performed {\it only on the
raw data of both $i_r$ and $i_n$}).

{\it (ii)} Calculate the transformation for $i_r$ to $i_n$ using equation
(3.5).

{\it (iii)} Generate a Difference Image (DI) by subtracting the transformed
$i_r$ and $i_n$ (note that at this point we have introduced a resampling
noise penalty due to the necessity of interpolating that data).

{\it (iv)} To the extent to which the residuals in the DI are statistically
significant those measurements are no longer considered, reducing the number 
of equations to be solved for from $N^2$ to $N'$. This statistical significance
test must be made relative to one of the two images, since the DI contains
structure due to the noise of sources above sky, even in the ideal case of
perfect subtraction.

{\it (v)} The total number of elements which $k$ now can be solved for is 
given by equation (3.7) with $N^2$ replaced by $N'$. Thus $k$ is recalculated
with these measurements.

{\it (vi)} The $N^2-N'$ bad pixels are interpolated over in both the {\it raw}
images $i_r$ and $i_n$ at the geometric location ${\bf x}$ in the former 
and ${\bf x'}$ in the latter image using the $N'$ good pixels.

{\it (vii)} Steps (ii) to (v) are repeated using the ``cleaned'' images
from step (vi) until all remaining measurements are statistically 
insignificant (step iv). Thus we iterate no further beyond step (v) and
have solved for $k$ to within the four limitations above.

\section{IMAGE DECONVOLUTION}

We now consider how to improve the spatial resolution of individual
images and then a combined image.

\subsection{Algorithm for deconvolving images to the spatial resolution
of the $i_r$ image}

{\it (i)} Take the FT of the ``cleaned'' $i_n$ raw image to yield $I_n$.

{\it (ii)} Take the FT of the kernel, $k_n$, determined for $i_n$ to yield
$K_n$.

{\it (iii)} Invert $K_n$ to $K_n^{-1}$.

{\it (iv)} Derive an image $(i_r)_n$ which is the FT of ($K_n^{-1} I_n$),
{\it i.e.,} the equivalent measurement of the $n$'th image at the spatial
resolution of the $i_r$ image.

To test whether $K$ has been solved for to within the limits of measurement
error the significance of the residuals in the DI of $i_r$ minus $(i_r)_n$
can be tested. However, to do this properly
we must geometrically and photometrically
transform the image simultaneously (using equation 3.5), 
thereby introducing some systematic
noise. Nevertheless these effects can be made very small.

\subsection{Algorithm for deconvolving images to approach the 
diffraction limited spatial resolution of the detector}

Now imagine we have a library of many images taken under many observing
different 
conditions, but we have deconvolved (or convolved) all the images to a
common spatial resolution using the steps outlined above. We wish to 
use these to try to achieve diffraction limited spatial resolution. One
of the biggest hurdles to overcome is that many images are heavily 
undersampled compared with the diffraction limit. Here we suggest a 
simple method to overcome this effect.

The angular size, $\theta$, of the minimum of flux between the centre of a 
PSF and the first diffraction order for a mirror 
or lens is given by the expression,

$$ \theta = B {\lambda \over {d}}, \eqno (3.8) $$

where $\lambda$ is the wavelength of the photons being measured and $d$ is
the physical size of detector. For a circular mirror or lens the constant
$B$ has a value of roughly 1.22. Within this ring we {\it cannot} distinguish 
between two point sources (represented by delta functions) and an extended
source.

Unfortunately detectors do not sample $\theta$ with infinitely 
discrete detecting elements. We can, however, extrapolate our images
to arbitrarily high resolution and use the information contained in 
$n$ deconvolved $(i_r)_n$ images taken at $n$ ${\bf x'}$ discrete locations to 
compensate for this effect. A high signal-to-noise (S/N) combined image, $R$, 
can be made by selecting those subpixels that correspond to locations
in ${\bf x}$ 
that suffer a minimal resampling noise penalty in the transformation
from ${\bf x'}$ to ${\bf x}$ 
({\it e.g.,} Mukai 1990) (in addition to corresponding to
original image $i$ pixels that have passed a DI statistical 
{\it in}significance 
test). An optimal choice must be made between the degree of subpixel sampling 
and the attainment of the best S/N for each subpixel in the final combined 
image in ${\bf x}$ coordinates for the $n$ images considered.

By averaging together our measurements at the ${\bf x}$
pixel or subpixel location 
of the $(i_r)_n$ image (or extrapolated image) and appropriately weighting 
the measurements with the inverse variance of the noise associated with each 
pixel, we can derive an $R$ of optimum S/N which can suppress some systematic
effects associated with the detector such as the variation of the quantum 
efficiency across an individual detector pixel as well as the resampling noise
and other sources of systematic noise associated with each subpixel for the 
$(i_r)_n$ images comprising $R$. 

The intrinsic spatially varying PSF $({\bf x})$ needs to be determined 
for an $R$ image. In astronomical images 
this can be done with standard algorithms such as 
DAOPHOT or DoPHOT which are described in detail in 
Stetson 1987 and Schechter, Mateo \& Saha 1993 respectively. Rather 
than detail the algorithms here we point out that after sources are
detected on the image and an estimate for the PSF is made at location
${\bf x}$ the sources must be subtracted from the image to within 
the measurement errors in order for the best possible estimate for the
PSF at $({\bf x})$ to be made. Again this is a similar iterative procedure 
involving detection, PSF estimation and source subtraction 
until all sources are removed to within the limits of measurement.
Once we have made an estimate of 
PSF $({\bf x})$ we may proceed with deconvolution of the image $R$ at
that ${\bf x}$ location to allow us to approach a diffraction-limited
image to within the limits of our measurement error ({\it e.g.,} Lucy 1992).

\section{SUMMARY}

Given the above arguments, the ideal imaging system comprises simply 
a re-imaging optic which brings to focus (although not necessarily a
perfect focus) a beam of photons onto a detector.
In the case where the re-imaging optic is a mirror (thereby suffering 
from minimal achromatic abberation) this detector is best removed from 
the incoming photon path by tilting the
mirror to bring the image to a focus off-axis. This removes
the strongly complicating effects on the final PSF and matching convolution 
kernel of diffraction due to other optical elements, but does 
introduce strong spatially dependent 
abberations in the PSF. Image distortion due to
this tilt can be reduced with a compensating tilt of the detector in the
the focal plane of the imager.
Diffraction-limited images can then be obtained to within the limits of 
measurement error by an iterative process of image differencing and 
deconvolution for a large database of images. It should be made clear that
in both a difference image and a deconvolved image we have not in any
way improved the signal-to-noise of a detection (and may have paid a 
penalty due to ``imperfect'' image processing) so 
we are still motived to 
increase our signal-to-noise ratio of our final combined and deconvolved 
image by ensuring the tightest possible focus of a point source on the 
detector in an original image. 

It is interesting to note that this re-imaging optic may comprise a 
gravitating body, which acts as an achromatic (but spherically
abberrated) lens.

A. B. T. would like to thank many friends and collegues in the development
of this paper, but especially the following: Arlin Crotts, Al Diercks,
Andy Gould and Chris Stubbs. A. B. T. is a member of
the Center for Particle Astrophysics which is supported in part by the Office 
of Science and Technology Centers if NSF under cooperative agreement 
AST 88-09616.

\end{document}